\documentclass[aps,pre,final,twocolumn,groupedaddress,showpacs]{revtex4}
\usepackage{amssymb,amsfonts,amsmath}
\usepackage{epsfig}
\usepackage{graphicx}
\begin{document}
\title{Crystallization of ion clouds in octupole traps: structural
  transitions, core melting, and scaling laws}



\author{F. Calvo}\email[]{fcalvo@lasim.univ-lyon1.fr}
\affiliation{LASIM, Universit\'e Claude Bernard Lyon 1 and UMR 5579,
CNRS, 43 Bd du 11 Novembre 1918, F69622 Villeurbanne Cedex, France}
\author{C. Champenois}
\affiliation{PIIM, UMR 6633 Universit\'e de Provence and CNRS, Campus
Universitaire de Saint-J\'er\^ome C21, F13397 Marseille cedex 20,
France}
\author{E. Yurtsever}
\affiliation{Ko\c c University, Rumelifeneriyolu, Sariyer, Istanbul
34450, Turkey}

\date{\today}

\begin{abstract}
The stable structures and melting properties of ion clouds in
isotropic octupole traps are investigated using a combination of
semi-analytical and numerical models, with a particular emphasis at
finite size scaling effects. Small-size clouds are found to be hollow
and arranged in shells corresponding approximately to the solutions of
the Thomson problem. The shell structure is lost in clusters
containing more than a few thousands of ions, the inner parts of the
cloud becoming soft and amorphous. While melting is triggered in the
core shells, the melting temperature unexpectedly follows the rule
expected for three-dimensional dense particles, with a depression
scaling linearly with the inverse radius.
\end{abstract}



\maketitle

\section{Introduction}
\label{sec:intro}

The storage of cold ions in electrostatic and magnetic fields (Penning
traps) or in radio-frequency electric fields (Paul traps) has become
possible with the advent of laser cooling techniques
\cite{neuhauser78,wineland78,itano98,mitchell98}. For few-particle
systems in quadrupole traps, this has lead to many applications
ranging from high resolution spectroscopy and optical frequency
standards \cite{rosenband08,chwalla09} to quantum information
\cite{kielpinski00,fsk03,leibfried03b} and tests on the possible
variations of the fundamental constants \cite{bize03,fortier07}. While
these effective harmonic traps allow focussing the ions at the center,
a greater number of ions with reduced rf driven motion can be stored
in higher-order confinements, with possible uses as microwave clocks
for deep space navigation \cite{prestage06} or to control cold
chemical reactions \cite{otto08} of astrophysical relevance.

Large samples of ions also offer a practical realization of the
classical many-body Coulomb problem with interesting collective
properties such as phase transitions. Crystallization of ion clouds
has been observed in quadrupole traps
\cite{diedrich87b,drewsen98,mortensen06}, in Penning traps
\cite{gilbert,itano98}, and theoretically studied by several groups
\cite{rafac91,totsuji02}. There, ion clouds crystallize into
well-defined shells at small sizes \cite{rafac91}, and then into the
bcc Wigner crystal in samples exceeding about $10^4$ ions
\cite{totsuji02}. Melting can be experimentally triggered by varying
the trap parameters \cite{hornekaer01}, and proceeds by separate
radial and orientational mechanisms in small clusters
\cite{drocco03}. Schiffer \cite{schiffer02} has reported from
molecular dynamics (MD) simulations that the large clouds melt from
the surface, resulting in a melting temperature that varies linearly
with inverse cluster radius.

In comparison, the properties of cold ion clouds in higher-order traps
are much less documented, despite specific investigations for octupole
traps \cite{walz94,okada07}. Using various semi-analytical and
numerical methods, we show in this article that the stable structures
of ion clouds in octupole traps are generally made of multiple
distinct shells only when the number of ions does not exceed a few
thousands. Above this approximate size, layering is progressively lost
except in the outermost regions of the cloud. We also found that
these clusters melt from the core and, quite unexpectedly, exhibit a
depression in the melting point that scales linearly with the cloud
radius.

In the next section, the model is described and the structural
properties of clusters are investigated. In Sec.~\ref{sec:temp}, the
finite temperature aspects are covered, and some concluding remarks
are finally given in Sec.~\ref{sec:ccl}.

\section{Model and static properties}
\label{sec:mod}

The system we are investigating consists of $N$ identical ions with
charge $q$ and mass $m$, trapped in an isotropic octopole trap. We
denote by $\Omega$ the radiofrequency of the electric field and by ${\cal
  E}({\bf r})$ its amplitude at position ${\bf r}$. We assume that the
adiabatic approximation holds and that the macromotion is driven by
the so-called pseudopotential $\Phi_{\rm T}({\bf r})$ as
\begin{equation}
\Phi_{\rm T}({\bf r}) = \frac{q^2}{4m\Omega^2}[{\cal E}({\bf r})]^2.
\label{eq:phitrap}
\end{equation}
We further assume that the rf-driven motion that superposes to the
macromotion can be neglected for the present purposes. For simplicity
of the following analysis, the electric field derives from a purely
radial three-dimensional octupole potential ${\cal E}=-{\rm grad} V$
with $V=V_0\times r^4$, $r$ being the distance from the trap center
and $V_0$ a constant. The pseudopotential then scales as $r^6$ and the
potential energy felt by an assembly of ions can thus be written as
\begin{eqnarray}
\Phi(\{{\bf r}_i\})&=& \Phi_{\rm T}(\{ {\bf r}_i \}) + \Phi_{\rm C}(\{
 {\bf r}_i \}), \nonumber \\
&=& A\sum_i r_i^6 + B\sum_{i<j} 1/r_{ij},
\label{eq:phitot}
\end{eqnarray}
where we have denoted by $r_i$ and $r_{ij}$ the distances of ion $i$
to the center of the trap and to particle $j$, respectively. The
constants $A$ and $B$ in the previous equation can be removed by
scaling of the quantities ${\bf r}\to \widetilde {\bf r}=\gamma {\bf
r}$ and $\Phi\to \widetilde \Phi=\Phi/\gamma B$, with
$\gamma=(A/B)^{1/7}$. In the following, reduced units are thus chosen
for both distances and energies, which amounts to using $A=B=1$ in
Eq.~(\ref{eq:phitot}).

We have first located the structures that globally minimize
Eq.~(\ref{eq:phitot}) at fixed size $N\leq 200$, employing the
basin-hopping Monte Carlo method \cite{wales97} successfully used in
previous related work \cite{calvo07}. The ions arrange into a single
spherical thin shell for $N<109$, and two shells above this size. This
shell structure can be fruitfully used to predict the minima for
larger clouds, assuming that the radial density is written as a sum
over $M$ concentric shells of zero thickness. Such a shell model was
initially developed for the quadrupole trap \cite{hasse91} and
improved for intra-shell correlations \cite{tsuruta93}. The energy to
be minimized, $\Phi_{\rm shell}$, is a function of the radii $\{
R_i\}$ of the shells, each shell $i$ carrying $N_i$ ions (plus one
possible ion at the center, $N_0$):
\begin{equation}
\Phi_{\rm shell}(\{ R_i,N_i\}) = \sum_{i=1}^M \frac{E_{\rm C}^{\rm
    intra}(N_i)}{R_i} + \frac{N_i}{R_i}\left(R_i^7 + \sum_{j<i} N_j \right).
\label{eq:phirini}
\end{equation}
In the above equation, we have denoted by $E_{\rm C}^{\rm intra}(N)$
the intra-shell Coulomb energy in which the particles lie on the unit
sphere. This energy is minimized at the solutions of the Thomson
problem \cite{thomson1904}, which have been tabulated up to rather
large sizes \cite{wales06}
\begin{equation}
E_{\rm C}^{\rm intra}(N) = E_{\rm Th}(N).
\label{eq:eth}
\end{equation}
Alternatively, for large numbers of ions, an asymptotic expression can
be substituted for the intra-shell Coulomb energy \cite{cioslowski08}
as
\begin{equation}
E_{\rm C}^{\rm intra}(N) = \frac{N(N-\alpha\sqrt{N})}{2},
\label{eq:easympt}
\end{equation}
where the $\alpha\sqrt{N}$ contribution accounts for intra-shell
correlations (neglecting correlations in a mean-field approach would
yield $N(N-1)/2$). Following the recent results of Cioslowski and
Grzebielucha \cite{cioslowski08} the parameter $\alpha$ was taken as
1.10610 to account for correlations. The above energy $\Phi_{\rm
shell}(\{R_i,N_i\})$ can be exactly minimized, resulting in the
expression for the radius of shell $i$
\begin{equation}
R_i = \left[\frac{1}{6}\left(\frac{E_{\rm C}^{\rm intra}(N_i)}{N_i}
+\sum_{j<i}N_j\right) \right]^{1/7}.
\label{eq:rinj}
\end{equation}
The remaining minimization of $\Phi_{\rm shell}$ with respect to the
$\{ N_i\}$ must then be carried out numerically under the constraint
of a fixed total number of ions.

When the tabulated optimal Thomson energies are used in place for
$E_{\rm C}^{\rm intra}(N)$, the minimization problem is variational
and provides rigorous upper bounds to the exact values. If the
asymptotic expression of Eq.~(\ref{eq:easympt}) is employed instead,
energies lower than the numerically exact minima may be reached due to
the approximate nature of this asymptotic form. The optimized
energies, outer radius, and shell arrangements obtained from Monte
Carlo minimization are listed in Table \ref{table} for selected
cluster sizes. In this table, the predictions of the shell models in
which the intra-shell energies are either taken from the tabulated
Thomson minima \cite{wales06} (discrete model) or from the asymptotic
expression (continuous model) are also given.

\begin{table*}[htb]
\begin{tabular}{r||r|c|c||r|c|c||r|c|c}
\hline
 & \multicolumn{3}{|c||}{Minimization} & \multicolumn{3}{|c}{Thomson
 shell model} & \multicolumn{3}{|c}{Asymptotic shell model}
 \\ \hline Size & Energy & Outer radius & Arrangement & Energy & Outer
 radius & Arrangement & Energy & Outer radius & Arrangement \\
\hline
10 &   41.624 & 0.917 & (10) &   41.624 & 0.917 & (10) &   41.399 & 0.916 & (10) \\
20 &  170.363 & 1.033 & (20) &  170.363 & 1.033 & (20) &  170.026 & 1.033 & (20) \\
30 &  380.045 & 1.104 & (30) &  380.045 & 1.104 & (30) &  379.611 & 1.104 & (30) \\
40 &  666.975 & 1.156 & (40) &  666.975 & 1.156 & (40) &  666.467 & 1.156 & (40) \\
50 & 1028.596 & 1.197 & (50) & 1028.596 & 1.197 & (50) & 1027.998 & 1.197 & (50) \\
60 & 1462.912 & 1.231 & (60) & 1462.912 & 1.231 & (60) & 1462.210 & 1.231 & (60) \\
70 & 1968.278 & 1.261 & (70) & 1968.278 & 1.261 & (70) & 1967.484 & 1.261 & (70) \\
80 & 2543.311 & 1.287 & (80) & 2543.311 & 1.287 & (80) & 2542.467 & 1.287 & (80) \\
90 & 3186.982 & 1.310 & (90) & 3186.983 & 1.310 & (90) & 3185.995 & 1.310 & (90) \\
100& 3898.102 & 1.331 & (100) & 3898.103 & 1.331 & (100) & 3897.051 & 1.331 & (100) \\
110& 4675.652 & 1.355 & (2,108) & 4675.858 & 1.355 & (2,108) & 4674.619 & 1.355 & (2,108) \\
120& 5518.467 & 1.375 & (4,116) & 5518.839 & 1.376 & (4,116) & 5517.415 & 1.374 & (3,117) \\
130& 6425.708 & 1.395 & (6,124) & 6426.152 & 1.396 & (6,124) & 6424.708 & 1.396 & (6,124) \\
140& 7396.731 & 1.413 & (8,132) & 7397.271 & 1.414 & (8,132) & 7395.796 & 1.414 & (8,132) \\
150& 8430.934 & 1.433 & (12,138) & 8431.678 & 1.433 & (12,138) & 8430.054 & 1.432 & (11,139) \\
160& 9527.713 & 1.448 & (14,146) & 9528.564 & 1.449 & (14,146) & 9526.851 & 1.447 & (13,147) \\
170& 10686.607 & 1.464 & (17,153) & 10687.544 & 1.464 & (17,153) & 10685.664 & 1.463 & (16,154) \\
180& 11906.937 & 1.479 & (20,160) & 11907.934 & 1.477 & (18,162) & 11905.975 & 1.478 & (19,161) \\
190& 13188.284 & 1.492 & (22,168) & 13189.315 & 1.492 & (22,168) & 13187.303 & 1.492 & (22,168) \\
200& 14530.226 & 1.506 & (26,174) & 14531.342 & 1.504 & (24,176) & 14529.198 & 1.506 & (25,175) \\
\hline
\end{tabular}
\caption{Lowest energy $E_N$, and outer shell radius $R_N$ found for
  ion clusters in the octupole trap, as obtained from Monte Carlo
  global minimization (left columns), from the discrete (Thomson) or
  continuous (asymptotic) shell models (central and right columns,
  respectively). The ion arrangements into shells predicted by the
  three methods are also indicated.}
\label{table}
\end{table*}

Energies, shell radius and ion arrangements agree very well between
the three methods, the shell models sometimes producing differences of
$\pm 1$ ion in the shells for the larger clusters containing 180 ions
or more. Energies and radii obtained from the tabulated Thomson minima
are in very good agreement with the exact results until two shells are
formed above 100 ions. The agreement is not as good when the asymptotic
form for the Thomson energies is used, the energy being slightly
underestimated. This underestimation suggests that the value of the
parameter $\alpha$ used in the shell model, which was taken from
extrapolations of the Thomson model to the large sizes regime
\cite{cioslowski08}, may be slightly excessive at small sizes. The
greater validity of the current value of $\alpha$ to large sizes is
also indicated by the relative error between the shell energy and the
Monte Carlo data, which for the systems considered in
Table~\ref{table} decreases from 0.5\% to less than 0.01\% as the
number of ions increases from 10 to 200. Despite this systematic
error, we generally find that the exact energy obtained by
minimization lies inbetween the predictions of the two shell models,
and that the outer radius is correctly reproduced (within 0.2\%) by
both models.

At this point, it is important to stress that the agreement between
the shell models and global optimization is essentially due to the
correct account of correlations, which are implicitly included in the
Thomson energies $E_{\rm Th}(N)$ or explicitly in the asymptotic
expression of Eq~(\ref{eq:easympt}) through the term $\delta
N=\alpha\sqrt{N}$. For comparison, the mean-field treatment predicts
that the structure of the 100-ion cluster would have 9 shells.

\begin{figure}[tb]
  \centerline{\includegraphics[width=8cm]{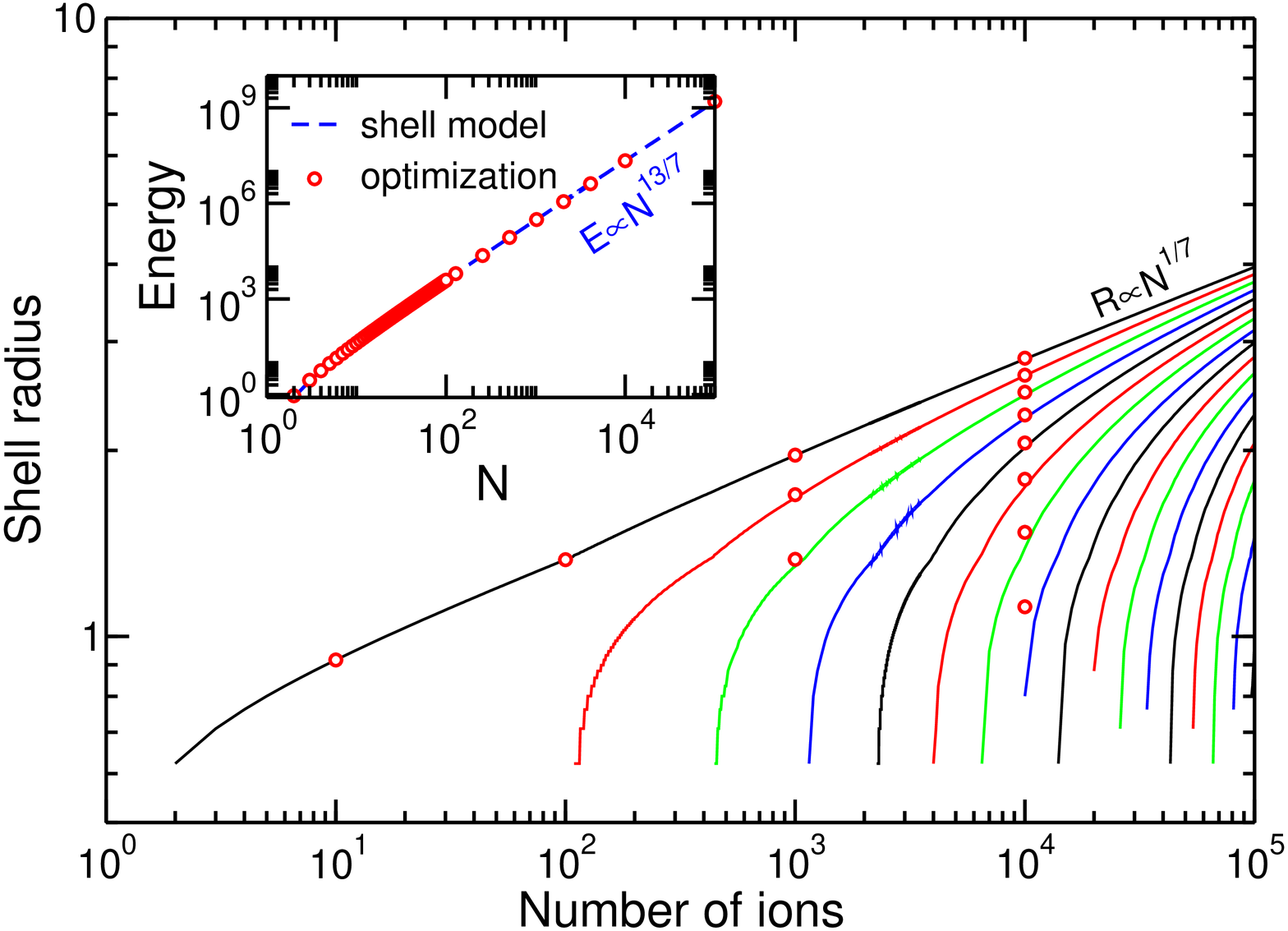}}
  \caption{(Color online) Shell radii versus cluster size, as predicted
    from the correlated shell model (solid lines; different curves
    correspond to successive shells) and from global optimization (open
    circles). The largest shell radius follows a $N^{1/7}$ scaling
    law. Inset: lowest energy versus cluster size, as obtained from the
    correlated shell model (dashed line) and from global optimization
    (open circles).}
  \label{fig:shell}
\end{figure}

The asymptotic shell model can be further exploited in the larger
sizes regime $10^3\leq N\leq 10^5$, where the optimal Thomson energies
are not systematically available. The shell radii predicted by this
model are represented as a function of the number of ions in
Fig.~\ref{fig:shell}. The average radii obtained from globally
optimized structures, superimposed for selected sizes $N=10^k$,
$k=1$--4, correctly match Eq.~(\ref{eq:rinj}) until about 1000 ions
are reached, larger clouds showing clear deviations in the inner
shells. However, the cloud radius (outermost shell) and the minimum
energy are both accurately reproduced. One prediction of the
continuous shell model is the size at which new shells appear. In
contrast with the quadrupole case, where shells are essentially added
over an existing core \cite{rafac91,totsuji02}, shells for octupole
clusters grow both on the outside and on the inside. The shell
capacitance may then be defined as the maximum number of ions that a
shell can sustain, and above which a new shell appears. The continuous
shell model predicts the onset of new shells at $N=109$, $N=442$,
$N=1129$, $N=2264$, $N=3992$, $N=6466$, $N=9709$, $N=13967$, and
$N=19249$. The model also predicts that the $10^5$-ion cluster should
have 17 shells but, as will be seen below, structural minimization
yields a more contrasted picture. These transitions occur at much
larger sizes than in the quadrupole case \cite{rafac91,totsuji02},
where for instance the 13- and 58-ion clusters adopt the (1,12) and
(1,12,45) shell structures, respectively \cite{beekman99}.

Up to 108 ions, the most stable configurations are the same as the
Thomson minima \cite{thomson1904,wales06}, within some minor radial
relaxations. The Thomson shell model also performs rather well for
predicting structures of clusters containing two shells. This
agreement suggests to use the known Thomson minima for optimizing the
atomic structure of multishell systems as well, as long as all shells
remain very thin. Such an idea is also supported by a recent
investigation by Cioslowski on a modified but similar Thomson model
\cite{cioslowski09}. We have guided Monte Carlo global minimization by
combining the results of the shell model with the available solutions
to the Thomson problem \cite{wales06}. In this approach, the
continuous shell model is used to predict the optimal number of layers
and the individual numbers of ions per layer (the discrete shell model
could also be used). The coordinates of the ions in each layer are
then scaled from the corresponding solution to the Thomson problem to
have a radius given by the predictions of the shell model, and the
only remaining degrees of freedom are two Euler angles
$\{\theta_i,\phi_i\}$ for each layer. The new Monte Carlo optimization
consists thus in first locating the low-energy regions in these angles
space, and to locally minimize the resulting structures by relaxing
all ionic positions. This method was found to perform very
satisfactorily with respect to brute force basin-hopping minimization
for several sizes in the range $N=100$--1000, leading to low-energy
structures often identical, or higher but by only a few $10^{-3}$
percents.

Provided that correlations are included in the evaluation of $E_{\rm
C}^{\rm intra}(N)$, the above results show that the combined
shell+Monte Carlo optimization should be especially useful for large
systems, for which successful basin-hopping runs would be prohibitive,
and this even allowed us to explore clusters containing up to $N=10^5$
ions. At large sizes, Fig.~\ref{fig:shell} shows that the cloud radius
$R$ and its minimum energy $\Phi$ scale as $R\propto N^{1/7}$ and
$\Phi \propto N^{13/7}$, respectively. Both scaling laws readily
follow from a simple cold fluid approximation, which is expected to be
valid for large sizes \cite{dubin99,cc09}. The maximum radius $R$ of
the cloud naturally relates to the total number $N$ of ions through a
$n(r)\propto r^4$ radial density. For the energy $\Phi$, the Coulomb
and trapping components are related to each other through the virial
theorem as $\Phi_{\rm C} = 6\Phi_{\rm T}$ at any local minimum
configuration where the gradient vanishes. Hence $\Phi$ is
proportional to the trapping energy only, which is simply integrated
over the radial density $n(r)$ as $\Phi=7\Phi_{\rm T}\propto
R^{13}\propto N^{13/7}$.

\begin{figure}[htb]
  \centerline{\includegraphics[width=8cm]{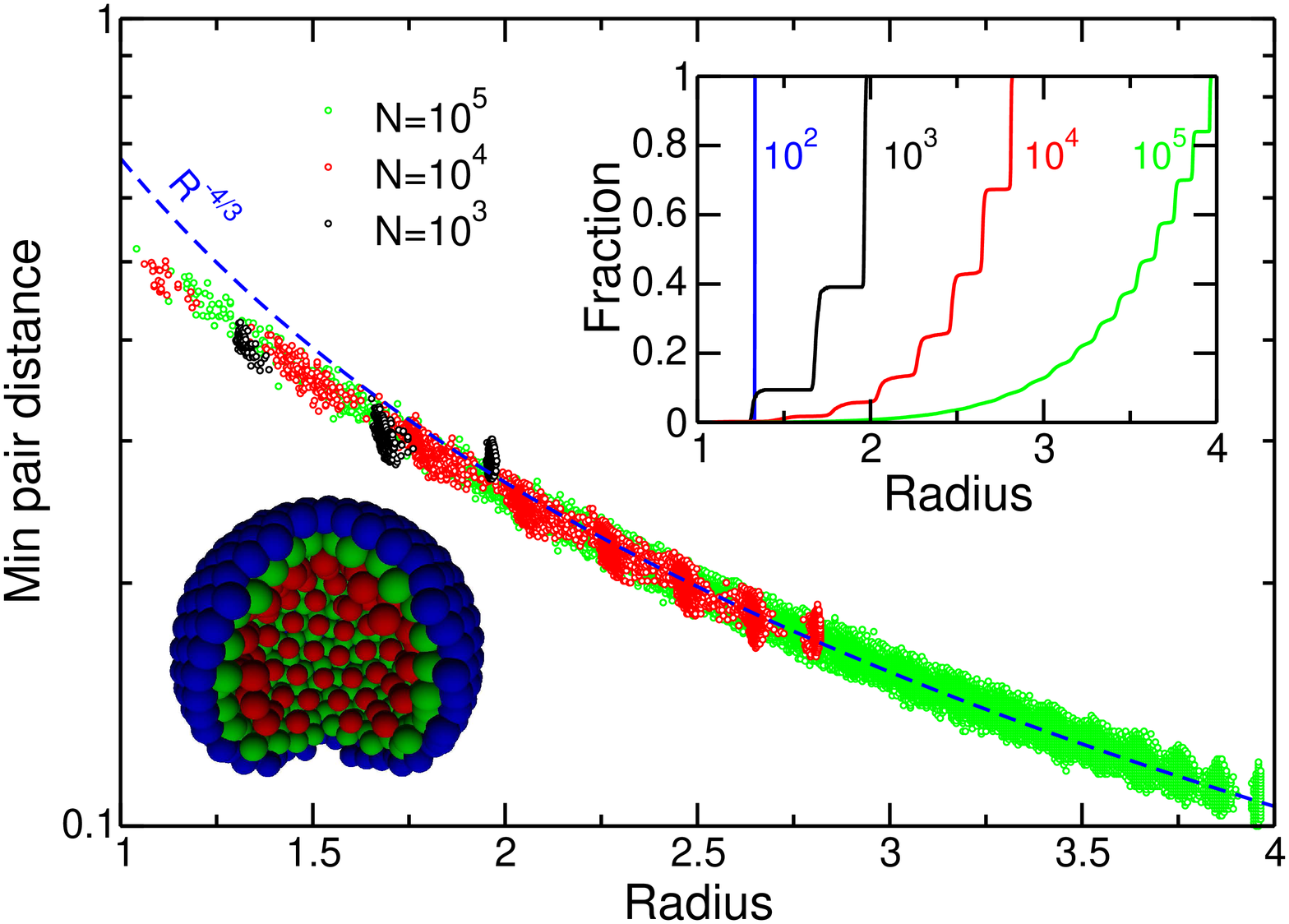}}
  \caption{(Color online) Minimum pair distance versus radial distance
    in stable structures containing 10$^3$--10$^5$ ions. Inset: fraction
    of ions inside a sphere of given radius. The most stable
    configuration of the 1000-ion cluster is also depicted, with the
    front quarter removed.}
  \label{fig:density}
\end{figure}

The validity of the cold fluid theory is better manifested on the
radial effective density profile. From the stable ionic configurations
obtained for the sizes $N=10^k$, $k=3$--5, we have calculated the
minimum pair distance $r_{ij}^{\rm min}$ between a given ion $i$ and
all other ions $j$, which is related to the local density $n(r)$
through $r_{ij}^{\rm min}\sim n^{-1/3}$. Fig.~\ref{fig:density} shows
the correlation between this quantity and the radial distance $r_i$ of
ion $i$ for the three sizes. The $r^{-4/3}$ dependence highlighted in
Fig.~\ref{fig:density} thus confirms that the cold fluid theory holds
increasingly well for these ion clouds, deviations being more
noticeable in the less populated (but more fluctuating) inner regions.
The most striking feature of Fig.~\ref{fig:density} is the broadening
of the shells when the number of ions increases from $10^3$ to $10^4$,
especially noticeable for the inner shells. Above a few thousands of
ions (an estimate based on mixed shell/Thomson optimization yields
$N^*\sim 4000$), this broadening is sufficient for some shells to
overlap into a more continuous radial distribution of ions below the
outermost layer, which is the only one to remain thin, even for the
largest size considered, $N=10^5$. This mixed continuous/discrete
behavior is better seen on the accumulated fraction $\chi(r)$ of ions
inside a sphere of radius $r$. The variations of $\chi$ with
increasing $r$, depicted as an inset in Fig.~\ref{fig:density},
exhibit spectacular changes as the size reaches $10^4$. The sharp
steps found in small clouds or at large radii are characteristic of
new shells, but are progressively softened in the inner regions of the
$10^4$-ion system and even replaced by a nearly continuous profile in
the largest cluster. In an octupole trap, crystallization should then
be understood as the formation of an outermost thin layer with a
softer, decreasingly dense but thick inner layer. Surface effects,
which play a major role in reducing the melting point of ion clouds in
harmonic potentials \cite{schiffer02}, could thus have a very
different influence in the case of the octupole trap.

\section{Finite temperature properties}
\label{sec:temp}

\begin{figure}[htb]
  \centerline{\includegraphics[width=8cm]{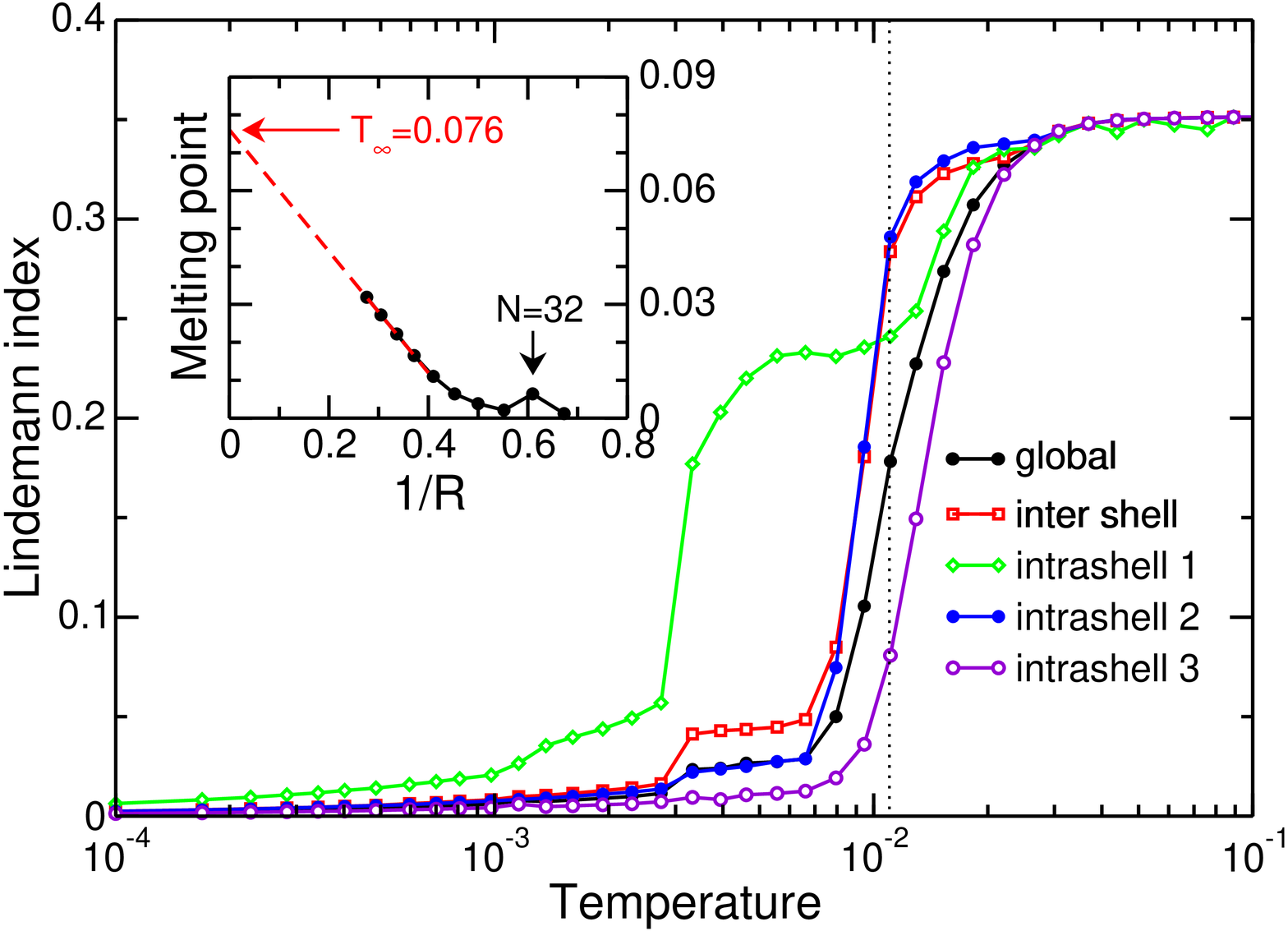}}
  \caption{(Color online) Relative bond length fluctuation indices
    obtained for the 512-ion cluster from molecular dynamics
    simulations. The dotted vertical line marks the onset of the melting
    temperature for this cluster ($T_m\simeq 1.1\times 10^{-2}$). Inset:
    dependence of the melting temperature with inverse cluster radius,
    and extrapolation to the infinitely large limit based on results
    obtained for $N\geq 512$.}
  \label{fig:delta}
\end{figure}

Whereas the previous section considered the stable structures and
static issues, we now discuss the thermodynamical and dynamical
behavior of selected ion clusters in octupole traps. Classical MD
simulations have been carried out at various energies to compute
several thermodynamical and structural observables. Order/disorder
transitions have been monitored using the root mean square bond length
fluctuation or Lindemann index $\delta$, as well as the
particle-resolved diffusion constant $D_i$ from the integrated
velocity autocorrelation function. The rms bond length fluctuation
index has been computed for collective parts of the clouds
(intra-shell indices) or from pairs of ions belonging to different
shells in the stable structure (inter-shell index). The variations of
the shell-resolved Lindemann indices with the kinetic temperature are
shown in Fig.~\ref{fig:delta} for the 512-ion system made of 3
shells. This size is small enough for the inner shells to be well
defined and not overlapping with each other.
All Lindemann indices exhibit smooth variations at low and high
temperatures, and one sharp increase above $\delta>0.15$ in a narrow
temperature range, allowing an accurate estimation of the
corresponding melting temperature. From Fig.~\ref{fig:delta}, the
global melting of the 512-ion system can therefore be located near
$T_m\simeq 1.1\times 10^{-2}$. However, the additional Lindemann
indices clearly reveal that the interior shell exhibits some
preliminary softening already above $T\simeq 10^{-3}$ before fully
melting near $T_m^{(1)} \simeq 3\times 10^{-3}$. Melting of this inner
shell impacts the distance fluctuations within the other shells, as
seen from the slight increase in the corresponding Lindemann
indices. However, these shells clearly undergo their own distinct
melting transitions at $T_m^{(2)} \simeq 8.5\times 10^{-3}$ (second
shell) and $T_m^{(3)}\simeq 1.4\times 10^{-2}$ (outermost shell). The
intershell Lindemann index follows similar variations as the
intrashell index of the intermediate, second shell. This agreement is
not fortuitous: melting of the second shell occurs while the interior
shell is already disordered, and the present results show that these
two shells tend to mix, to some extent, while the outermost layer
remains rigid.

\begin{figure}[htb]
  \centerline{\includegraphics[width=8cm]{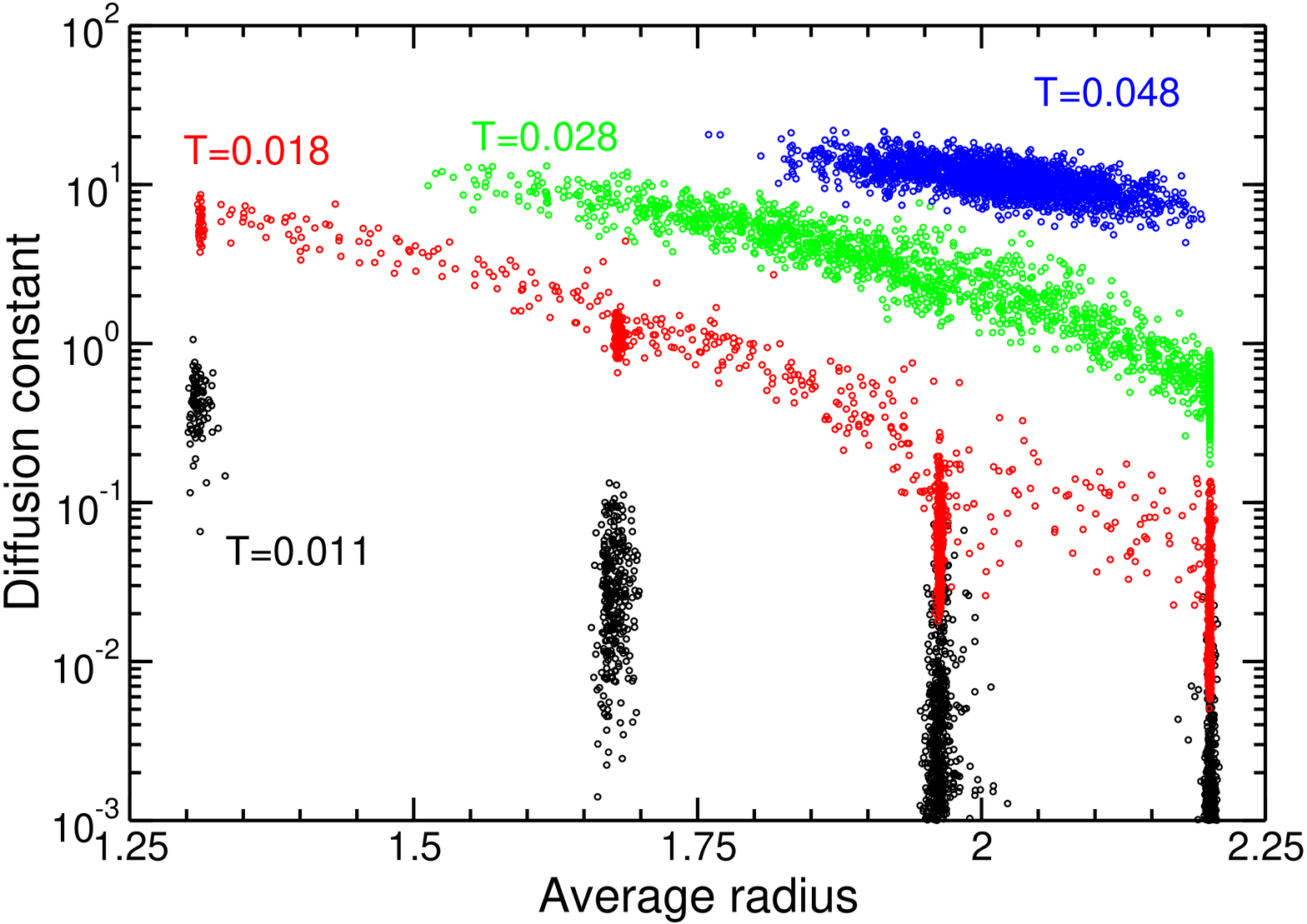}}
  \caption{(Color online) Ion diffusion constant as a function of its
    time average radius, in the 2048-ion cluster at four temperatures.}
  \label{fig:atomr}
\end{figure}

The melting mechanisms can be analysed in more details by looking at
the ions motion at various temperatures. The correlations between the
average radial distance $\langle r_i\rangle$ and the diffusion
constant $D_i$ are shown in Fig.~\ref{fig:atomr} for the 2048-ion
system at four characteristic temperatures. For this system, the
melting point obtained from the variations of the global Lindemann
index lies near $T_m\simeq 0.023$. At $T=0.011$, the four shells are
clearly seen as narrow vertical spots with low values for $D_i$ at
radii close to 1.32, 1.68, 1.97, and 2.20. The intershell motion and
occasional hops of ions between neighboring shells are seen at
$T=0.018$, they are associated with a much higher diffusion
constant. Note that $D_i$ exhibit a steady decrease with the radial
distance, in agreement with the previously found softening of internal
layers. At $T=0.028$ the global diffusivity is also higher, and the
shell structure seems essentially lost. A similar trend is found at
the highest temperature considered here, $T=0.048$, with a single
broad spot centered around a radius of 2.02.
The peculiar core melting effect is a consequence of the lower density
of the inner shells, and is similar in this respect to the general
surface melting process in solid state and nanoscale materials
\cite{tartaglino05}. At first sight, this phenomenon may preclude from
unambiguously defining the melting temperature of the entire system
since, strictly speaking, the system is not yet fully melted when the
global Lindemann index barely exceeds 0.15. However, because the
relative number of ions in the external regions grows with size as
$N^{1/7}$ according to the cold fluid model, the definition of the
global index $\delta$ should reflect more and more closely the value
of the outermost layer. The melting point $T_m$ was thus defined for
all sizes as the temperature at which $\delta$ exceeds 0.15. For the
present clouds in octupole traps, the variations of $T_m$ have been
represented in the inset of Fig.~\ref{fig:delta} as a function of
inverse cluster radius. Not surprisingly, fluctuations are seen at
small sizes, and the 32-ion system ($1/R \simeq 0.61$) seems extra
resistant to melting relative to its neighboring sizes. Since the 31-
and 33-ion systems have a much lower melting point, the special
stability of the 32-ion cluster indicates a magic character, further
confirmed by its high icosahedral symmetry. As $N$ reaches 256 ($1/R
\simeq 0.45$), $T_m$ increases linearly as the inverse radius
decreases. This linear depression, though not anticipated for these
highly heterogeneous, core-melted systems, allows some straightforward
extrapolating to the infinite size limit $1/R\to 0$, leading to
$T_\infty \simeq 0.076$. Even though it is hard to figure out what a
cloud confined in an octupole trap would physically represent at the
bulk limit, the present simulation results should be very useful for
estimating melting temperatures in system sizes of the order of $10^5$
or more, as experimentally studied in other groups \cite{drewsen98}.

\section{Concluding remarks}
\label{sec:ccl}

Despite their obvious differences, ion clouds in octupole and
quadrupole confinements share several remarkable features. Firstly,
the onset of the transition from a multishell structure to the bcc
Wigner crystal in quadrupole trap was estimated to be around $10^4$
ions \cite{totsuji02}. In octupole traps the shell structure becomes
blurred above a few thousands of ions. Secondly, for both cases, the
depression in the melting point with respect to the bulk limit follows
a linear scaling with inverse cluster radius \cite{schiffer02}.
Thirdly, in small systems, a clear differential melting between the
inner and outer shells seems to take place in both confinements. These
similarities may well extend past the specific octupole trap, hold
also in higher-order traps, and even be universal. As the exponent
$p=6$ in Eq.~(\ref{eq:phitot}) takes higher integer values, the stable
shell structure should be qualitatively preserved, only with fewer
shells. Therefore the transition to a softened core should be delayed
as $p$ increases, but one cannot exclude that the same melting
mechanisms will remain, and in particular that the melting temperature
will display linear variations with inverse radius.

Because ion clouds in octupole traps remain poorly studied, the
present work could be extended along many lines. Keeping the isotropic
case as a model of a more realistic three-dimensional trap, it would
be interesting to examine more specifically dynamical properties
involving the vibrations and normal modes \cite{dubin91,tellez97} or
the transitions between regular and chaotic regimes
\cite{gaspard03,yurtsever05}. Extension of analytical
models \cite{cioslowski08,cioslowski09} developed for the quadrupole
case to higher-order traps would also be useful.

Finally, at the price of introducing additional parameters in the
model, one natural step beyond the present work would be to look at
clusters in linear octupole traps, such as those discussed by Okada
and coworkers \cite{okada07}. In these systems, confinement along the
symmetry axis is harmonic in nature, whereas confinement
perpendicular to the axis is octupolar. The stable structures are
hollow tubes of ions \cite{okada07}, and it is unclear how the
combined contributions of the harmonic and octupolar confinements
will determine the cluster properties. Even aware of these
complications, the multishell structure and core-melted phase
predicted here, as well as the scaling laws connecting size, radius,
and melting temperature, should all become amenable to experimental
comparison in the near future. May such measurements also help us
understanding the collective properties of these exotic states of
Coulombic matter.

\begin{acknowledgments}
This work was supported by ANR-JCJC-0053-01. E.Y. would like to
acknowledge the support of the Turkish Academy of Sciences. Molecular
dynamics simulations have been carried out at the P\^ole Scientifique
de Mod\'elisation Num\'erique, which is gratefully acknowledged.
\end{acknowledgments}


\begin{thebibliography}{30}
\expandafter\ifx\csname natexlab\endcsname\relax\def\natexlab#1{#1}\fi
\expandafter\ifx\csname bibnamefont\endcsname\relax
  \def\bibnamefont#1{#1}\fi
\expandafter\ifx\csname bibfnamefont\endcsname\relax
  \def\bibfnamefont#1{#1}\fi
\expandafter\ifx\csname citenamefont\endcsname\relax
  \def\citenamefont#1{#1}\fi
\expandafter\ifx\csname url\endcsname\relax
  \def\url#1{\texttt{#1}}\fi
\expandafter\ifx\csname urlprefix\endcsname\relax\def\urlprefix{URL }\fi
\providecommand{\bibinfo}[2]{#2}
\providecommand{\eprint}[2][]{\url{#2}}


\bibitem{neuhauser78} W. Neuhauser, M. Hohenstatt, P. E. Toschek and
  H. G. Dehmelt, Phys. Rev. Lett. {\bf 41}, 233 (1978).

\bibitem{wineland78} D. J. Wineland, R. E.  Drullinger and
  F. L. Walls, Phys. Rev. Lett. {\bf 40}, 1639 (1978).

\bibitem{itano98} W. M. Itano, J. J. Bollinger, J. N. Tan,
  B. Jelenkovi\'c, X.-P. Huang, and D. J. Wineland, Science {\bf 279},
  686-689 (1998).

\bibitem{mitchell98} T. B. Mitchell, J. J. Bollinger, D. H. E. Dubin,
  X.-P. Huang, W. M. Itano, and R. H. Baughman, Science {\bf 282},
  1290-1293 (1998).

\bibitem[{\citenamefont{Rosenband et~al.}(2008)\citenamefont{Rosenband, Hume,
  Schmidt, Chou, Brusch, Lorini, Oskay, Drullinger, Fortier, Stalnaker
  et~al.}}]{rosenband08}
\bibinfo{author}{\bibfnamefont{T.}~\bibnamefont{Rosenband}},
  \bibinfo{author}{\bibfnamefont{D.~B.} \bibnamefont{Hume}},
  \bibinfo{author}{\bibfnamefont{P.~O.} \bibnamefont{Schmidt}},
  \bibinfo{author}{\bibfnamefont{C.~W.} \bibnamefont{Chou}},
  \bibinfo{author}{\bibfnamefont{A.}~\bibnamefont{Brusch}},
  \bibinfo{author}{\bibfnamefont{L.}~\bibnamefont{Lorini}},
  \bibinfo{author}{\bibfnamefont{W.~H.} \bibnamefont{Oskay}},
  \bibinfo{author}{\bibfnamefont{R.~E.} \bibnamefont{Drullinger}},
  \bibinfo{author}{\bibfnamefont{T.~M.} \bibnamefont{Fortier}},
  \bibinfo{author}{\bibfnamefont{J.~E.} \bibnamefont{Stalnaker}},
  \bibnamefont{et~al.}, \bibinfo{journal}{Science}
  \textbf{\bibinfo{volume}{319}}, \bibinfo{pages}{1808} (\bibinfo{year}{2008}).

\bibitem[{\citenamefont{Chwalla et~al.}(2009)\citenamefont{Chwalla, Benhelm,
  Kim, Kirchmair, Monz, Riebe, Schindler, Villar, H\"{a}nsel, Roos
  et~al.}}]{chwalla09}
\bibinfo{author}{\bibfnamefont{M.}~\bibnamefont{Chwalla}},
  \bibinfo{author}{\bibfnamefont{J.}~\bibnamefont{Benhelm}},
  \bibinfo{author}{\bibfnamefont{K.}~\bibnamefont{Kim}},
  \bibinfo{author}{\bibfnamefont{G.}~\bibnamefont{Kirchmair}},
  \bibinfo{author}{\bibfnamefont{T.}~\bibnamefont{Monz}},
  \bibinfo{author}{\bibfnamefont{M.}~\bibnamefont{Riebe}},
  \bibinfo{author}{\bibfnamefont{P.}~\bibnamefont{Schindler}},
  \bibinfo{author}{\bibfnamefont{A.~S.} \bibnamefont{Villar}},
  \bibinfo{author}{\bibfnamefont{W.}~\bibnamefont{H\"{a}nsel}},
  \bibinfo{author}{\bibfnamefont{C.~F.} \bibnamefont{Roos}},
  \bibnamefont{et~al.}, \bibinfo{journal}{Phys. Rev. Lett.}
  \textbf{\bibinfo{volume}{102}}, \bibinfo{pages}{023002}
  (\bibinfo{year}{2009}).

\bibitem[{\citenamefont{Kielpinski et~al.}(2000)\citenamefont{Kielpinski, King,
  Myatt, Sackett, Turchette, Itano, Monroe, Wineland, and
  Zurek}}]{kielpinski00}
\bibinfo{author}{\bibfnamefont{D.}~\bibnamefont{Kielpinski}},
  \bibinfo{author}{\bibfnamefont{B.~E.} \bibnamefont{King}},
  \bibinfo{author}{\bibfnamefont{C.~J.} \bibnamefont{Myatt}},
  \bibinfo{author}{\bibfnamefont{C.~A.} \bibnamefont{Sackett}},
  \bibinfo{author}{\bibfnamefont{Q.~A.} \bibnamefont{Turchette}},
  \bibinfo{author}{\bibfnamefont{W.~M.} \bibnamefont{Itano}},
  \bibinfo{author}{\bibfnamefont{C.}~\bibnamefont{Monroe}},
  \bibinfo{author}{\bibfnamefont{D.~J.} \bibnamefont{Wineland}},
  \bibnamefont{and} \bibinfo{author}{\bibfnamefont{W.~H.} \bibnamefont{Zurek}},
  \bibinfo{journal}{Phys. Rev. A} \textbf{\bibinfo{volume}{61}},
  \bibinfo{pages}{032310} (\bibinfo{year}{2000}).

\bibitem[{\citenamefont{Schmidt-Kaler et~al.}(2003)\citenamefont{Schmidt-Kaler,
  H\"affner, Riebe, Gulde, Lancaster, Deutschle, Becher, Roos, Eschner, and
  Blatt}}]{fsk03}
\bibinfo{author}{\bibfnamefont{F.}~\bibnamefont{Schmidt-Kaler}},
  \bibinfo{author}{\bibfnamefont{H.}~\bibnamefont{H\"affner}},
  \bibinfo{author}{\bibfnamefont{M.}~\bibnamefont{Riebe}},
  \bibinfo{author}{\bibfnamefont{S.}~\bibnamefont{Gulde}},
  \bibinfo{author}{\bibfnamefont{G.~P.~T.} \bibnamefont{Lancaster}},
  \bibinfo{author}{\bibfnamefont{T.}~\bibnamefont{Deutschle}},
  \bibinfo{author}{\bibfnamefont{C.}~\bibnamefont{Becher}},
  \bibinfo{author}{\bibfnamefont{C.}~\bibnamefont{Roos}},
  \bibinfo{author}{\bibfnamefont{J.}~\bibnamefont{Eschner}}, \bibnamefont{and}
  \bibinfo{author}{\bibfnamefont{R.}~\bibnamefont{Blatt}},
  \bibinfo{journal}{Nature (London)} \textbf{\bibinfo{volume}{422}},
  \bibinfo{pages}{408} (\bibinfo{year}{2003}).

\bibitem[{\citenamefont{Leibfried et~al.}(2003)\citenamefont{Leibfried,
  DeMarco, Meyer, Lucas, Barrett, Britton, Itano, Jelenkovic, Langer, Rosenband
  et~al.}}]{leibfried03b}
\bibinfo{author}{\bibfnamefont{D.}~\bibnamefont{Leibfried}},
  \bibinfo{author}{\bibfnamefont{B.}~\bibnamefont{DeMarco}},
  \bibinfo{author}{\bibfnamefont{V.}~\bibnamefont{Meyer}},
  \bibinfo{author}{\bibfnamefont{D.}~\bibnamefont{Lucas}},
  \bibinfo{author}{\bibfnamefont{M.}~\bibnamefont{Barrett}},
  \bibinfo{author}{\bibfnamefont{J.}~\bibnamefont{Britton}},
  \bibinfo{author}{\bibfnamefont{W.~M.} \bibnamefont{Itano}},
  \bibinfo{author}{\bibfnamefont{B.}~\bibnamefont{Jelenkovic}},
  \bibinfo{author}{\bibfnamefont{C.}~\bibnamefont{Langer}},
  \bibinfo{author}{\bibfnamefont{T.}~\bibnamefont{Rosenband}},
  \bibnamefont{et~al.}, \bibinfo{journal}{Nature (London)}
  \textbf{\bibinfo{volume}{422}}, \bibinfo{pages}{412} (\bibinfo{year}{2003}).

\bibitem[{\citenamefont{Bize et~al.}(2003)\citenamefont{Bize, Diddams, Tanaka,
  Tanner, Oskay, Drullinger, Parker, Heavner, Jefferts, Hollberg
  et~al.}}]{bize03}
\bibinfo{author}{\bibfnamefont{S.}~\bibnamefont{Bize}},
  \bibinfo{author}{\bibfnamefont{S.}~\bibnamefont{Diddams}},
  \bibinfo{author}{\bibfnamefont{U.}~\bibnamefont{Tanaka}},
  \bibinfo{author}{\bibfnamefont{C.}~\bibnamefont{Tanner}},
  \bibinfo{author}{\bibfnamefont{W.}~\bibnamefont{Oskay}},
  \bibinfo{author}{\bibfnamefont{R.}~\bibnamefont{Drullinger}},
  \bibinfo{author}{\bibfnamefont{T.}~\bibnamefont{Parker}},
  \bibinfo{author}{\bibfnamefont{T.}~\bibnamefont{Heavner}},
  \bibinfo{author}{\bibfnamefont{S.}~\bibnamefont{Jefferts}},
  \bibinfo{author}{\bibfnamefont{L.}~\bibnamefont{Hollberg}},
  \bibnamefont{et~al.}, \bibinfo{journal}{Phys. Rev. Lett.}
  \textbf{\bibinfo{volume}{90}}, \bibinfo{pages}{150802}
  (\bibinfo{year}{2003}).

\bibitem[{\citenamefont{Fortier et~al.}(2007)\citenamefont{Fortier, Ashby,
  Bergquist, Delaney, Diddams, Heavner, Hollberg, Itano, Jefferts, Kim
  et~al.}}]{fortier07}
\bibinfo{author}{\bibfnamefont{T.~M.} \bibnamefont{Fortier}},
  \bibinfo{author}{\bibfnamefont{N.}~\bibnamefont{Ashby}},
  \bibinfo{author}{\bibfnamefont{J.~C.} \bibnamefont{Bergquist}},
  \bibinfo{author}{\bibfnamefont{M.~J.} \bibnamefont{Delaney}},
  \bibinfo{author}{\bibfnamefont{S.~A.} \bibnamefont{Diddams}},
  \bibinfo{author}{\bibfnamefont{T.~P.} \bibnamefont{Heavner}},
  \bibinfo{author}{\bibfnamefont{L.}~\bibnamefont{Hollberg}},
  \bibinfo{author}{\bibfnamefont{W.~M.} \bibnamefont{Itano}},
  \bibinfo{author}{\bibfnamefont{S.~R.} \bibnamefont{Jefferts}},
  \bibinfo{author}{\bibfnamefont{K.}~\bibnamefont{Kim}}, \bibnamefont{et~al.},
  \bibinfo{journal}{Phys. Rev. Lett.} \textbf{\bibinfo{volume}{98}},
  \bibinfo{pages}{070801} (\bibinfo{year}{2007}).

\bibitem[{\citenamefont{Prestage et~al.}(2006)\citenamefont{Prestage, Chung,
  Le, Lim, and Maleki}}]{prestage06}
\bibinfo{author}{\bibfnamefont{J.}~\bibnamefont{Prestage}},
  \bibinfo{author}{\bibfnamefont{S.}~\bibnamefont{Chung}},
  \bibinfo{author}{\bibfnamefont{T.}~\bibnamefont{Le}},
  \bibinfo{author}{\bibfnamefont{L.}~\bibnamefont{Lim}}, \bibnamefont{and}
  \bibinfo{author}{\bibfnamefont{L.}~\bibnamefont{Maleki}}, in
  \emph{\bibinfo{booktitle}{Proceedings of IEEE Int. Freq. Contr. Symp. Miami,
  jun. 5-7, 2006}} (\bibinfo{publisher}{IEEE, New York}, \bibinfo{year}{2006}).

\bibitem[{\citenamefont{Otto et~al.}(2008)\citenamefont{Otto, Mikosch, Trippel,
  Weidem\"{u}ller, and Wester}}]{otto08}
\bibinfo{author}{\bibfnamefont{R.}~\bibnamefont{Otto}},
  \bibinfo{author}{\bibfnamefont{J.}~\bibnamefont{Mikosch}},
  \bibinfo{author}{\bibfnamefont{S.}~\bibnamefont{Trippel}},
  \bibinfo{author}{\bibfnamefont{M.}~\bibnamefont{Weidem\"{u}ller}},
  \bibnamefont{and} \bibinfo{author}{\bibfnamefont{R.}~\bibnamefont{Wester}},
  \bibinfo{journal}{Phys. Rev. Lett.} \textbf{\bibinfo{volume}{101}},
  \bibinfo{pages}{063201} (\bibinfo{year}{2008}).

\bibitem[{\citenamefont{Diedrich et~al.}(1987)\citenamefont{Diedrich, Peik,
  Chen, Quint, and Walther}}]{diedrich87b}
\bibinfo{author}{\bibfnamefont{F.}~\bibnamefont{Diedrich}},
  \bibinfo{author}{\bibfnamefont{E.}~\bibnamefont{Peik}},
  \bibinfo{author}{\bibfnamefont{J.~M.} \bibnamefont{Chen}},
  \bibinfo{author}{\bibfnamefont{W.}~\bibnamefont{Quint}}, \bibnamefont{and}
  \bibinfo{author}{\bibfnamefont{H.}~\bibnamefont{Walther}},
  \bibinfo{journal}{Phys. Rev. Lett.} \textbf{\bibinfo{volume}{59}},
  \bibinfo{pages}{2931} (\bibinfo{year}{1987}).

\bibitem[{\citenamefont{Drewsen et~al.}(1998)\citenamefont{Drewsen, Brodersen,
  Hornek\ae{}r, Hangst, and Schiffer}}]{drewsen98}
\bibinfo{author}{\bibfnamefont{M.}~\bibnamefont{Drewsen}},
  \bibinfo{author}{\bibfnamefont{C.}~\bibnamefont{Brodersen}},
  \bibinfo{author}{\bibfnamefont{L.}~\bibnamefont{Hornek\ae{}r}},
  \bibinfo{author}{\bibfnamefont{J.~S.} \bibnamefont{Hangst}},
  \bibnamefont{and} \bibinfo{author}{\bibfnamefont{J.~P.}
  \bibnamefont{Schiffer}}, \bibinfo{journal}{Phys. Rev. Lett.}
  \textbf{\bibinfo{volume}{81}}, \bibinfo{pages}{2878} (\bibinfo{year}{1998}).

\bibitem[{\citenamefont{Mortensen et~al.}(2006)\citenamefont{Mortensen,
  Nielsen, Matthey, and Drewsen}}]{mortensen06}
\bibinfo{author}{\bibfnamefont{A.}~\bibnamefont{Mortensen}},
  \bibinfo{author}{\bibfnamefont{E.}~\bibnamefont{Nielsen}},
  \bibinfo{author}{\bibfnamefont{T.}~\bibnamefont{Matthey}}, \bibnamefont{and}
  \bibinfo{author}{\bibfnamefont{M.}~\bibnamefont{Drewsen}},
  \bibinfo{journal}{Phys. Rev. Lett.} \textbf{\bibinfo{volume}{96}},
  \bibinfo{eid}{103001} (\bibinfo{year}{2006}).

\bibitem{gilbert} S. L. Gilbert, J. J. Bollinger, and D. J. Wineland,
  Phys. Rev. Lett. {\bf 60}, 2022 (1988).

\bibitem[{\citenamefont{Rafac et~al.}(1991)\citenamefont{Rafac, Schiffer,
  Handst, Dubin, and Wales}}]{rafac91}
\bibinfo{author}{\bibfnamefont{R.}~\bibnamefont{Rafac}},
  \bibinfo{author}{\bibfnamefont{J.~P.} \bibnamefont{Schiffer}},
  \bibinfo{author}{\bibfnamefont{J.~S.} \bibnamefont{Handst}},
  \bibinfo{author}{\bibfnamefont{D.~H.} \bibnamefont{Dubin}}, \bibnamefont{and}
  \bibinfo{author}{\bibfnamefont{D.~J.} \bibnamefont{Wales}},
  \bibinfo{journal}{Proc. Nat. Acad. Sci. USA} \textbf{\bibinfo{volume}{88}},
  \bibinfo{pages}{483} (\bibinfo{year}{1991}).

\bibitem[{\citenamefont{Totsuji et~al.}(2002)\citenamefont{Totsuji, Kishimoto,
  Totsuji, and Tsuruta}}]{totsuji02}
\bibinfo{author}{\bibfnamefont{H.}~\bibnamefont{Totsuji}},
  \bibinfo{author}{\bibfnamefont{T.}~\bibnamefont{Kishimoto}},
  \bibinfo{author}{\bibfnamefont{C.}~\bibnamefont{Totsuji}}, \bibnamefont{and}
  \bibinfo{author}{\bibfnamefont{K.}~\bibnamefont{Tsuruta}},
  \bibinfo{journal}{Phys. Rev. Lett.} \textbf{\bibinfo{volume}{88}},
  \bibinfo{pages}{125002} (\bibinfo{year}{2002}).

\bibitem[{\citenamefont{Hornekaer et~al.}(2001)\citenamefont{Hornekaer,
  Kjaergaard, Thommesen, and Drewsen}}]{hornekaer01}
\bibinfo{author}{\bibfnamefont{L.}~\bibnamefont{Hornekaer}},
  \bibinfo{author}{\bibfnamefont{N.}~\bibnamefont{Kjaergaard}},
  \bibinfo{author}{\bibfnamefont{A.M.}~\bibnamefont{Thommesen}},
  \bibnamefont{and} \bibinfo{author}{\bibfnamefont{M.}~\bibnamefont{Drewsen}},
  \bibinfo{journal}{Phys. Rev. Lett.} \textbf{\bibinfo{volume}{86}},
  \bibinfo{pages}{1994} (\bibinfo{year}{2001}).

\bibitem[{\citenamefont{Drocco et~al.}(2003)\citenamefont{Drocco, Reichhardt,
  Reichhardt, and Jank\'o}}]{drocco03}
\bibinfo{author}{\bibfnamefont{J.~A.} \bibnamefont{Drocco}},
  \bibinfo{author}{\bibfnamefont{C.~J.} \bibnamefont{Olson Reichhardt}},
  \bibinfo{author}{\bibfnamefont{C.}~\bibnamefont{Reichhardt}},
  \bibnamefont{and} \bibinfo{author}{\bibfnamefont{B.}~\bibnamefont{Jank\'o}},
  \bibinfo{journal}{Phys. Rev. E} \textbf{\bibinfo{volume}{68}},
  \bibinfo{pages}{060401(R)} (\bibinfo{year}{2003}).

\bibitem[{\citenamefont{Schiffer}(2002)}]{schiffer02}
\bibinfo{author}{\bibfnamefont{J.~P.} \bibnamefont{Schiffer}},
  \bibinfo{journal}{Phys. Rev. Lett.} \textbf{\bibinfo{volume}{88}},
  \bibinfo{pages}{205003} (\bibinfo{year}{2002}).

\bibitem[{\citenamefont{Walz et~al.}(1994)\citenamefont{Walz, Siemers,
  Schubert, Neuhauser, Blatt, and Teloy}}]{walz94}
\bibinfo{author}{\bibfnamefont{J.}~\bibnamefont{Walz}},
  \bibinfo{author}{\bibfnamefont{I.}~\bibnamefont{Siemers}},
  \bibinfo{author}{\bibfnamefont{M.}~\bibnamefont{Schubert}},
  \bibinfo{author}{\bibfnamefont{W.}~\bibnamefont{Neuhauser}},
  \bibinfo{author}{\bibfnamefont{R.}~\bibnamefont{Blatt}}, \bibnamefont{and}
  \bibinfo{author}{\bibfnamefont{E.}~\bibnamefont{Teloy}},
  \bibinfo{journal}{Phys. Rev. A} \textbf{\bibinfo{volume}{50}},
  \bibinfo{pages}{4122} (\bibinfo{year}{1994}).

\bibitem[{\citenamefont{Okada et~al.}(2007)\citenamefont{Okada, Yasuda,
  Takayanagi, Wada, Schuessler, and Ohtani}}]{okada07}
\bibinfo{author}{\bibfnamefont{K.}~\bibnamefont{Okada}},
  \bibinfo{author}{\bibfnamefont{K.}~\bibnamefont{Yasuda}},
  \bibinfo{author}{\bibfnamefont{T.}~\bibnamefont{Takayanagi}},
  \bibinfo{author}{\bibfnamefont{M.}~\bibnamefont{Wada}},
  \bibinfo{author}{\bibfnamefont{H.~A.} \bibnamefont{Schuessler}},
  \bibnamefont{and} \bibinfo{author}{\bibfnamefont{S.}~\bibnamefont{Ohtani}},
  \bibinfo{journal}{Phys. Rev. A} \textbf{\bibinfo{volume}{75}},
  \bibinfo{eid}{033409} (\bibinfo{year}{2007}).

\bibitem[{\citenamefont{Wales and Doye}(1997)}]{wales97}
\bibinfo{author}{\bibfnamefont{D.~J.} \bibnamefont{Wales}} \bibnamefont{and}
  \bibinfo{author}{\bibfnamefont{J.~P.~K.} \bibnamefont{Doye}},
  \bibinfo{journal}{J. Phys. Chem. A} \textbf{\bibinfo{volume}{101}},
  \bibinfo{pages}{5111} (\bibinfo{year}{1997}).

\bibitem[{\citenamefont{Calvo and Yurtsever}(2007)}]{calvo07}
\bibinfo{author}{\bibfnamefont{F.}~\bibnamefont{Calvo}} \bibnamefont{and}
  \bibinfo{author}{\bibfnamefont{E.}~\bibnamefont{Yurtsever}},
  \bibinfo{journal}{Eur. Phys. J. D} \textbf{\bibinfo{volume}{44}},
  \bibinfo{pages}{81} (\bibinfo{year}{2007}).

\bibitem[{\citenamefont{Hasse and Avilov}(1991)}]{hasse91}
\bibinfo{author}{\bibfnamefont{R.~W.} \bibnamefont{Hasse}} \bibnamefont{and}
  \bibinfo{author}{\bibfnamefont{V.~V.} \bibnamefont{Avilov}},
  \bibinfo{journal}{Phys. Rev. A} \textbf{\bibinfo{volume}{44}},
  \bibinfo{pages}{4506} (\bibinfo{year}{1991}).

\bibitem[{\citenamefont{Tsuruta and Ichimaru}(1993)}]{tsuruta93}
\bibinfo{author}{\bibfnamefont{K.}~\bibnamefont{Tsuruta}} \bibnamefont{and}
  \bibinfo{author}{\bibfnamefont{S.}~\bibnamefont{Ichimaru}},
  \bibinfo{journal}{Phys. Rev. A} \textbf{\bibinfo{volume}{48}},
  \bibinfo{pages}{1339} (\bibinfo{year}{1993}).
  
  \bibitem[{\citenamefont{Thomson}(1904)}]{thomson1904}
\bibinfo{author}{\bibfnamefont{J.~J.} \bibnamefont{Thomson}},
  \bibinfo{journal}{Phil. Mag.} \textbf{\bibinfo{volume}{7}},
  \bibinfo{pages}{237} (\bibinfo{year}{1904}).
  
  \bibitem[{\citenamefont{Wales and Ulker}(2006)}]{wales06}
\bibinfo{author}{\bibfnamefont{D.~J.} \bibnamefont{Wales}} \bibnamefont{and}
  \bibinfo{author}{\bibfnamefont{S.}~\bibnamefont{Ulker}},
  \bibinfo{journal}{Phys. Rev.~B} \textbf{\bibinfo{volume}{74}},
  \bibinfo{pages}{212101} (\bibinfo{year}{2006}).


\bibitem[{\citenamefont{Cioslowski and Grzebielucha}(2008)}]{cioslowski08}
\bibinfo{author}{\bibfnamefont{J.}~\bibnamefont{Cioslowski}} \bibnamefont{and}
  \bibinfo{author}{\bibfnamefont{E.}~\bibnamefont{Grzebielucha}},
  \bibinfo{journal}{Phys. Rev.~E} \textbf{\bibinfo{volume}{78}},
  \bibinfo{pages}{026416} (\bibinfo{year}{2008}).

\bibitem{beekman99} R. A. Beekman, M. R. Roussel, and P. J. Wilson,
  Phys. Rev. A {\bf 59}, 503 (1999).




\bibitem{cioslowski09} J. Cioslowski, Phys. Rev. E {\bf 79}, 046405
  (2009).

\bibitem[{\citenamefont{Dubin and O\char39{}Neil}(1999)}]{dubin99}
\bibinfo{author}{\bibfnamefont{D.~H.~E.} \bibnamefont{Dubin}} \bibnamefont{and}
  \bibinfo{author}{\bibfnamefont{T.~M.} \bibnamefont{O\char39{}Neil}},
  \bibinfo{journal}{Rev. Mod. Phys.} \textbf{\bibinfo{volume}{71}},
  \bibinfo{pages}{87} (\bibinfo{year}{1999}).

\bibitem[{\citenamefont{Champenois}(2009)}]{cc09}
\bibinfo{author}{\bibfnamefont{C.}~\bibnamefont{Champenois}},
  \bibinfo{journal}{J. Phys. B: At. Mol. Opt. Phys.}
  \textbf{\bibinfo{volume}{42}}, \bibinfo{pages}{154002}
  (\bibinfo{year}{2009}).

\bibitem[{\citenamefont{Tartaglino et~al.}(2005)\citenamefont{Tartaglino,
  Zykova-Timan, Ercolessi, and Tosatti}}]{tartaglino05}
\bibinfo{author}{\bibfnamefont{U.}~\bibnamefont{Tartaglino}},
  \bibinfo{author}{\bibfnamefont{T.}~\bibnamefont{Zykova-Timan}},
  \bibinfo{author}{\bibfnamefont{F.}~\bibnamefont{Ercolessi}},
  \bibnamefont{and} \bibinfo{author}{\bibfnamefont{E.}~\bibnamefont{Tosatti}},
  \bibinfo{journal}{Phys. Rep.} \textbf{\bibinfo{volume}{411}},
  \bibinfo{pages}{191} (\bibinfo{year}{2005}).

\bibitem{dubin91} D. H. E. Dubin, Phys. Rev. Lett. {\bf 66},
  2076 (1991).

\bibitem{tellez97} G. T\'ellez, Phys. Rev. E {\bf 55}, 3400
  (1997).

\bibitem{gaspard03} P. Gaspard, Phys. Rev. E {\bf 68},
  056209 (2003).

\bibitem{yurtsever05} E. Yurtsever, F. Calvo, and D. J. Wales,
  Phys. Rev. E {\bf 72}, 026110 (2005).
\end{thebibliography}
\end{document}